\documentclass[journal]{IEEEtran}
\IEEEoverridecommandlockouts
\usepackage{lipsum}
\usepackage{cite}
\usepackage{amsmath,amssymb,amsfonts}
\usepackage{algorithmic}
\usepackage{graphicx}
\graphicspath{ {./images/} }
\usepackage{textcomp}
\usepackage{indentfirst}
\usepackage{caption}
\usepackage{hyperref}
\usepackage{soul}
\usepackage{xcolor}
\usepackage[left=3.5cm,right=2.5cm,top=2.5cm,includefoot,bottom=2.5cm]{geometry}  
\title{A Federated Learning Scheme for Neuro-developmental Disorders: Multi-Aspect ASD Detection}

\begin{document}
\author{
    \IEEEauthorblockN{Hala Shamseddine $^\dag$, Safa Otoum $^\ddag$, Azzam Mourad $^\dag$$^\star$}\\
    
    \IEEEauthorblockA{$^\dag$ Cyber Security Systems and Applied AI Research Center, Department of CSM, Lebanese American University, Lebanon}
    
    \IEEEauthorblockA{$^\ddag$ College of Technological Innovation (CTI), Zayed University, United Arab Emirates}
    
    \IEEEauthorblockA{$^\star$ Division of Science, New York University, Abu Dhabi, United Arab Emirates}
    
    \IEEEauthorblockA{
    {
    \href{mailto:hala.shamseddine@lau.edu.lb}{hala.shamseddine}@lau.edu.lb,
    \href{mailto:safa.otoum@zu.ac.ae}{Safa.Otoum}@zu.ac.ae,
    \href{mailto:azzam.mourad@lau.edu.lb}{azzam.mourad}@lau.edu.lb
    }
    }
    }

\renewcommand{\labelenumii}{\arabic{enumi}.\arabic{enumii}}

\maketitle
\begin{abstract}
Autism Spectrum Disorder (ASD) is a neuro-developmental syndrome resulting from alterations in the embryological brain before birth. This disorder distinguishes its patients by special socially restricted and repetitive behavior in addition to specific behavioral traits. Hence, this would possibly deteriorate their social behavior among other individuals, as well as their overall interaction within their community. Moreover, medical research has proved that ASD also affects the facial characteristics of its patients, making the syndrome recognizable from distinctive signs within an individual's face. Given that as a motivation behind our work, we propose a novel privacy-preserving federated learning scheme to predict ASD in a certain individual based on their behavioral and facial features, embedding a merging process of both data features through facial feature extraction while respecting patient data privacy. After training behavioral and facial image data on federated machine learning models, promising results are achieved, with 70\% accuracy for the prediction of ASD according to behavioral traits in a federated learning environment, and a 62\% accuracy is reached for the prediction of ASD given an image of the patient's face. Then, we test the behavior of regular as well as federated ML on our merged data, behavioral and facial, where a 65\% accuracy is achieved with the regular logistic regression model and 63\% accuracy with the federated learning model.

\end{abstract}


\section{\textbf{Introduction}}
To start with, Autism Spectrum disorder (ASD) could be defined as a complex neuro-developmental syndrome \cite{comp} characterizing its patients by specific socially restricted and repetitive behavior. In other words, this disorder affects the social and behavioral traits of the patient, hence deteriorating their social behavior among other individuals, as well as their overall interaction within their community. According to \cite{facialpheno}, ASD could be a result of alterations in the embryological brain pre-birth, making it a brain-based disorder affecting patients. 

Furthermore, ASD patients are distinguished from non-ASD persons not only by unique social and behavioral characteristics, but also by distinct facial features, as demonstrated by \cite{deveval}.
According to the authors of \cite{craniofacial}, neuro-developmental disorders such as ASD can cause craniofacial abnormalities, or atypical facial features. 
For instance, \cite{sexspecific} and \cite{hyper} prove that female and male individuals with higher levels of autistic traits are associated with less feminine and masculine facial structures, respectively. Furthermore, facial measurements taken in \cite{clinicalsub} form a superb autism bio-marker. As a result, according to past medical studies, ASD sufferers have distinct facial traits that identify them from normal individuals.

Some previous studies have proposed predicting whether a patient has ASD using Machine learning (ML), a sub-field of Artificial Intelligence (AI). ML is based on giving a computer program the ability to learn from data so that new predictions are made \cite{whatsml}. This learning process \cite{supervised} could be divided into supervised learning, when data is labelled, or unsupervised learning, when the trainer learns patterns on unlabeled data. Machine learning applications \cite{class} are majorly split into classification and regression, as the predicted value is either a discrete class/label or a continuous value, respectively. With the advancement of this science, a major application of ML has appeared in the healthcare \cite{sysrev} and medical fields \cite{rahman}, specifically in disease prediction \cite{corrigendum}. Predicting ASD in children could help in the early diagnosis of this syndrome, which may be made possible using ML. However, data used for training such a machine learner is medical data, thus sensitive and classified as private to patients. Hence, to preserve the privacy of patients, we propose using a federated learning-based privacy-preserving framework \cite{fedmlsurvey1} to predict whether a patient has ASD or not, based on behavioral as well as facial image data. 

Federated Learning is based on combining privacy with regular ML. Training on data by ML algorithms occurs on distributed sites, called clients, where raw data is not shared among clients \cite{fedmlsurvey2}, maintaining patients' privacy. On the other hand, each client shares their local ML model weights, after learning its bulk of data, with a global aggregator who averages model parameters of all learned models by clients into a global model used for prediction. To the best of our knowledge, none of the previous works in the literature have used federated learning for predicting ASD using behavioral or facial image data. We, therefore, provide a novel privacy-preserving federated learning scheme for predicting ASD based on behavioral data and facial data separately, and then propose a merging method, through feature extraction, for both data sets to predict ASD according to a combination of both behavioral and facial traits.

The main contributions of this work are summarized as follows:
\begin{itemize}
    \item Introducing a privacy-preserving federated learning model for predicting ASD based on behavioral and social characteristics of a patient.
    
    \item Proposing a federated learning model for the diagnosis of ASD based on a clear image of the patient's face.
    
    
    \item Through combining both behavioral and facial datasets via feature extraction methods, we devised a novel federated learning scheme for predicting ASD in patients from a behavioral and facial perspective.
    
    
   

\end{itemize}

The rest of the paper is organized as follows. In Section \ref{related}, we tackle some of the previous work in the literature. In Section \ref{proposed}, we illustrate the proposed FL scheme, discuss its components, and describe the model evaluation criteria. Next, in section \ref{merge}, we present the combined scheme of both behavioral and facial datasets under study and introduce the multi-aspect ASD prediction federated learning model. In Section \ref{results}, we introduce the datasets used for training the machine learners, the FL Setup, and the experimental results. In Section \ref{dis}, we discuss the results in comparison with previous methods and evaluate the performance of the proposed scheme. Finally, we conclude the paper in Section \ref{conc}.

\section{\textbf{Related Work}} \label{related}

Several fields have proved the efficiency of applying regular as well as federated ML \cite{fedml} to predict certain real-world answers. 
According to authors in \cite{sysrev}, a major application of ML would be in the field of medicine and healthcare, such as predicting certain human diseases, traits, or syndromes \cite{adhdsev}. Furthermore, in \cite{medimgfed}, the authors tackle the issue of medical data scarcity and sensitivity, by proposing an additional level of privacy protection in medical imaging data through the use of federated privacy-preserving artificial intelligence. In federated machine learning \cite{systematiclit}, data is stored locally at each distributed client, training on it locally without moving the data outside the client devices, thus preserving the data privacy.  

For instance, in \cite{heartdm} and \cite{heartml}, authors proposed several machine learning as well as data mining methodologies for the sake of predicting heart diseases in patients, given their seriousness and the importance of early diagnosis of such diseases. Moreover, authors in \cite{parkml} introduced machine learning models for predicting Parkinson's disease based on biomedical voice measurements of patients. To add more, medical image classification \cite{deepdiag} plays a vital role in the clinical diagnosis of remarkable diseases and thus helps in the treatment process through early discovery of illnesses and disorders. Then, authors in \cite{deepdiag} used deep learning methods to prove its effectiveness through disease prediction using neural network models.

Another effective main syndrome prediction using ML techniques would be Autism Spectrum Disorder (ASD), which is what we focus on in this paper.  
In \cite{therapy} and \cite{eye}, authors used ML classification models with 80\% accuracy to predict whether a child is autistic or not based on behavioral data collected automatically from tablet games and eye-tracking data captured from face-to-face conversations with individuals, respectively. Also, in \cite{earlydet}, eye-tracking information and image content was used to detect if an individual has ASD. Moreover, \cite{induction} proposed a new ML technique called Rules-ML that reveals autistic traits of cases while offering knowledge rules, achieving an accuracy higher than other regular ML strategies. Furthermore, \cite{comp} improved the measured accuracy in ASD classification, reaching a near 100\% accuracy using the Multi-Layer Perceptron ML model on behavioral specific collected data. 

According to \cite{multiple}, who used the same behavioral dataset as we aim at using in this paper, found that using random forests, decision trees, and support vector machine algorithms results in around 90\% accuracy, which improves with using feature selection methods. A remarkable dataset used in the literature was the ABIDE dataset, which consisted of MRI neuro-images in \cite{deepabide} and \cite{multisite}. Deep learning models were proposed in \cite{deepabide} to classify ASD vs. non-ASD patients, achieving a 70\% accuracy. Using the same dataset of neuro-images, federated learning was used to associate an MRI image with an autistic individual or not \cite{multisite}, preserving the privacy of screened patients. However, none of the previous studies mentioned in the literature, to the best of our knowledge, has used federated learning to predict ASD based on behavioral data and facial features.

\renewcommand\thesection{\Roman{section}}

\section{\textbf{Proposed Scheme}} \label{proposed}
In this section, we present the problem definition and scheme overview, including the evaluation metrics. 

\subsection{ \textbf{Problem Definition}}
Notable applications for ML have been recently demonstrated in the fields of healthcare and medicine, specifically for predicting whether a certain patient carries a specific disease or not. In this context, we tackle the research problem of predicting Autism Spectrum Disorder (ASD) based on a set of input features or traits, based on two data formats separately. We then test the same problem within a combined approach, representing the different aspects of the disorder: behavioral and facial.
    
First, given a set of behavioral features, including age and gender, in addition to behavior-specific traits, our federated learning model would predict whether the corresponding patient is autistic or not, having learned securely on the training data discussed in section \ref{datasets}. Moreover, the second part of our work, which includes another aspect of ASD for prediction, is based on the facial features of a patient. Given a clear image of their face, showing all of their facial sides and characteristics, our model would also return a classification of ASD for the respective patient, while preserving the privacy of data and images. Third, given a combination of these features, behavioral as well as facial landmarks, our model would be able to accurately classify whether the individual is autistic or not, by extracting facial distances from facial images and combining these features with behavioral ones.

\subsection{\textbf{Scheme Overview}}
\begin{figure}[h]
\centering
\includegraphics[scale=0.4]{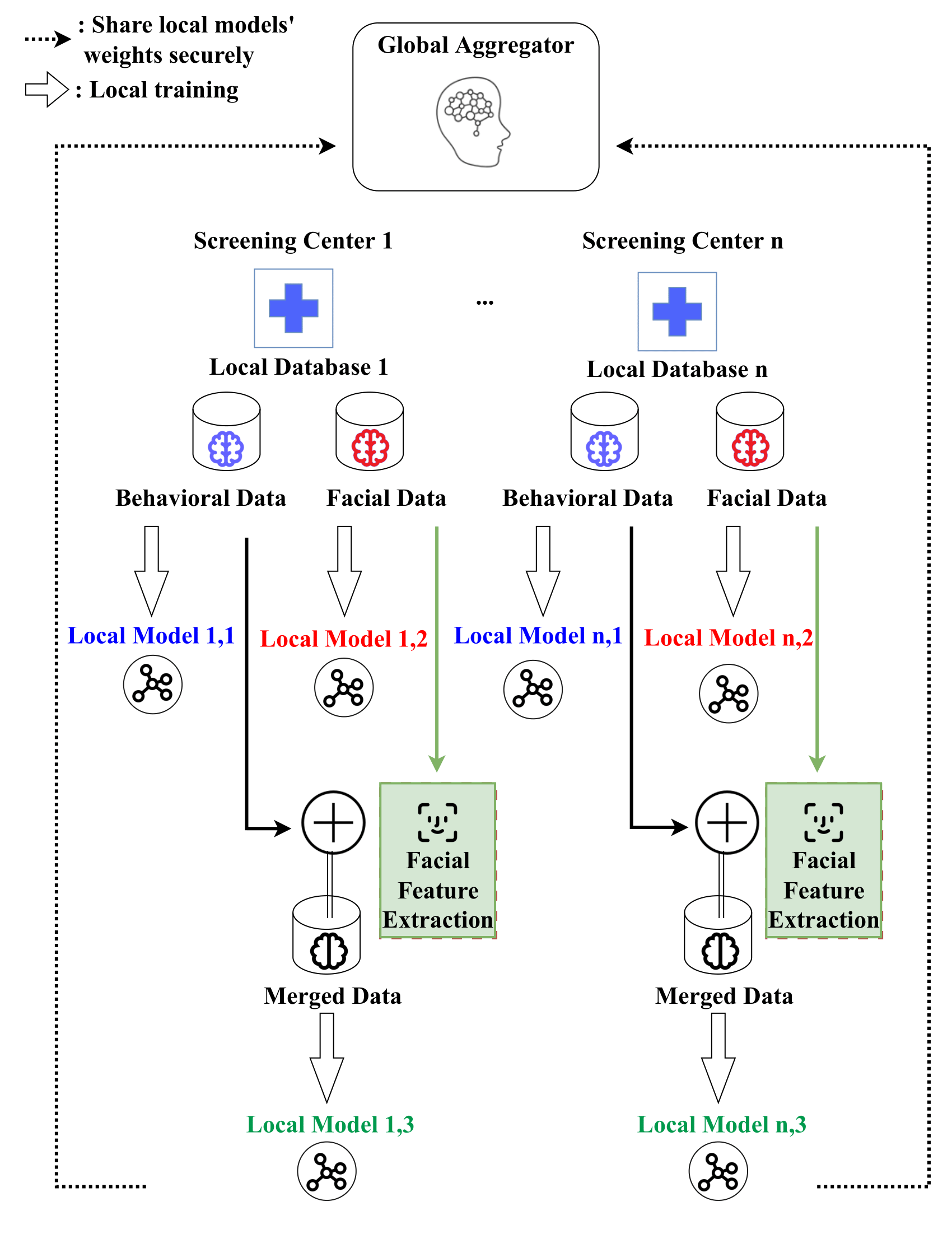}
\caption{Federated Machine Learning Training for ASD Prediction}
\captionsetup{justification=centering}
\label{fig:my_labelarch}
\end{figure}

Figure \ref{fig:my_labelarch} illustrates the architecture of our proposed scheme. As could be clearly noticed, we have two main entities: the ASD screening centers as well as the global aggregator.
First, given a set of $n$ ASD screening centers $Sc_1, Sc_2 ... Sc_n$, each conducting certain surveys on screening ASD patients, screening images for the same patients, or collecting any data related to the latter. These results and data instances are stored locally within the local database of every center, without any sharing with external entities, given the confidentiality and sensitivity of data at hand, and respecting patients' privacy. Therefore, every center has a local database $Ld_i$ that includes the data collected. In our case, data is split into two: the first local dataset contains behavioral data, and the second contains facial data.

Then, for the sake of training, every entity would train on its own data stored in its local database, $Ld_i$. For example, $Sc_1$ would train on $Ld_1$, $Sc_2$ on $Ld_2$, etc. Hence, every training process would result in a local ML model trained on the corresponding data in its local database, represented as:
\begin{equation}\label{firstequ}
    M_i = \sum_{i,j= 1}^{i,j= m} w_{i,j} . X_{i}   
\end{equation}
where $w_{1,1}, w_{1,2}, ... w_{i,m}$ are the $m$ local model weights and $X_1, X_2 ... X_m$ are $m$ model features.

To clarify, the screening center $Sc_1$, having a local database $Ld_1$, contains behavioral as well as facial data. First, local training on the behavioral data takes place, producing a local model $M_1$. Then, the client trains locally on the facial dataset within the same local database $Ld_1$, resulting in a local model $M_2$. Finally, facial feature extraction is done on the facial data in database $Ld_1$, and extracted facial features are then merged with behavioral data in dataset $Ld_1$ of the same client (more details are provided in Section \ref{merge}). This process results in a third local dataset consisting of merged data containing behavioral as well as facial features, on which the client also trains locally, producing a third local model  $M_3$. These steps are repeated by all the clients, from which three local models are obtained for each one: the first one from training on behavioral data, the second on facial data, and the third on merged data.

In order to achieve a global model capable of combining the ML results of all the centers into a single framework, a global aggregator (GA) is introduced. Once local training is done at each screening entity, and an $M_i$ local model is obtained at each $Sc_i$, the ML model weights $w_{1,1}, w_{1,2}, ... w_{i,m}$ are forwarded by the clients into the GA, without sharing the original patient data. Therefore, for example, in the screening center, $Sc_1$, weights of local models $M_1$, $M_2$, and $M_3$ corresponding to each dataset (behavioral, facial, or merged), are shared securely with the GA. Then, $Sc_2$ also shares the weights of its local models $M_1$, $M_2$, and $M_3$ with the GA, and similarly for the rest of the screening centers. 

Finally, the GA averages the local weights for each local model separately, producing 3 global models $GM$s: the first one for predicting ASD based on behavioral traits, the second based on facial traits, and the third based on a combination of behavioral and facial features.
In order to aggregate models' weights, averaging is done as follows:
\begin{equation}\label{secondequ}
GM = \sum_{i=1}^{i=m} h_i . X_i 
\end{equation}
with $h_i$, represented in \ref{Thirdequ}, being the averaged weight for each feature variable $X_i$: 
\begin{equation}\label{Thirdequ}
h_i = \frac {\sum_{i=1}^{m} w_{i}}{n}
\end{equation}
where n is the total number of clients. 




\subsection{ \textbf{Model Evaluation}}    
In order to evaluate each of the ML models used in our scheme, we measure each client's accuracy value. This metric represents how far the predicted answer is from the actual classification, which is later aggregated into a global accuracy demonstrating the overall model's performance. In the federated learning scheme, at each local training phase, data is split into 80\% training data and the remaining 20\% is kept for testing. Thus, once training is done, we measure the accuracy value by testing how many correct predictions are made in comparison with the real testing data having known labels or classes. 

A Binary Cross-Entropy loss is used for calculating this accuracy value at each client locally. Cross-entropy loss measures the performance of a classification model having an output with a probability value between 0 and 1. Its value increases while the predicted probability diverges from the actual label. As our classification is binary; we only have two classes, ASD and non-ASD. We choose this loss function as a suitable measure. Hence, the higher the number of correct predictions, the lower the cross-entropy loss, and the higher the obtained accuracy value $Acc_i$ is. 

The global accuracy $GAcc_i$ is calculated by averaging the local accuracy values $Acc_i$ of all local models trained by clients throughout an experiment, similar to the averaging method of local model weights previously presented: 
\begin{equation} \label{fourtheq}
    GAcc_i = \frac{\sum_{i= 1}^{i= n} Acc_i . Sd_i}{\sum_{i= 1}^{i= n} Sd_i}
\end{equation}
with n being the total number of clients, $Sd_i$ being the sample data size at client i, and $Acc_i$ being the local accuracy produced by each local model at a client i. 

\section{\textbf{Multi-Aspect ASD Prediction Approach}} \label{merge}

In this paper, we tackle two significant aspects of ASD: a behavioral aspect, affecting the social behavior of an autistic patient within their community, and a facial aspect affecting the physical features of a patient's face. We have discussed in previous sections how each of these aspects of ASD is used by machine learners. However, in this section, we discuss the process of merging both datasets, \cite{behvdataset} behavioral and \cite{facialdataset} facial image, motivated by testing the behavior of machine learning for the prediction of ASD depending on the combination of multiple human aspects (e.g. behavioral and facial). Then, given a combination of behavioral as well as facial features of a certain individual, our federated machine learning model would predict whether this individual has ASD or not, rather than using each of these aspects separately. 

\vspace{1mm}
\begin{enumerate}
    
    \vspace{3mm}
    \item \textbf{Data Generation} \\
Two datasets are used in this study: a behavioral dataset \cite{behvdataset} consisting of 610 instances, and a facial image dataset \cite{facialdataset} containing around 2900 images. Given that these two datasets do not have equal sizes, merging them is challenging. To tackle this issue, we built a random instance generator for behavioral data, producing instances with values in the ranges of those of real data. Then, we end up with around 2900 behavioral data instances, equally split among ASD and non-ASD classes, which is equal to the number of instances in the image dataset. Therefore, as the two datasets now have equal size, one behavioral data instance could be mapped to a facial image from the second dataset depending on their class.

    \item \textbf{Facial Feature Extraction} \\
Another problem faced in this experiment would be merging datasets of different types: CSV dataset \cite{behvdataset} and image dataset \cite{facialdataset}. As we previously mentioned, medical research studies in \cite{clinicalsub}, have proved that ASD affects the facial characteristics of its patients, differentiating them from non-autistic individuals. To prove that, authors in \cite{clinicalsub} measure facial distances after examining the facial landmarks of the patient's face in comparison to healthy patients. Hence, we rely on this work in order to merge both datasets by extracting certain features from each of the images in the dataset and mapping these facial features into one behavioral instance rather than mapping the whole image.

    \vspace{1mm}
     \begin{figure}[h]
        \centering
        \includegraphics[scale=0.3]{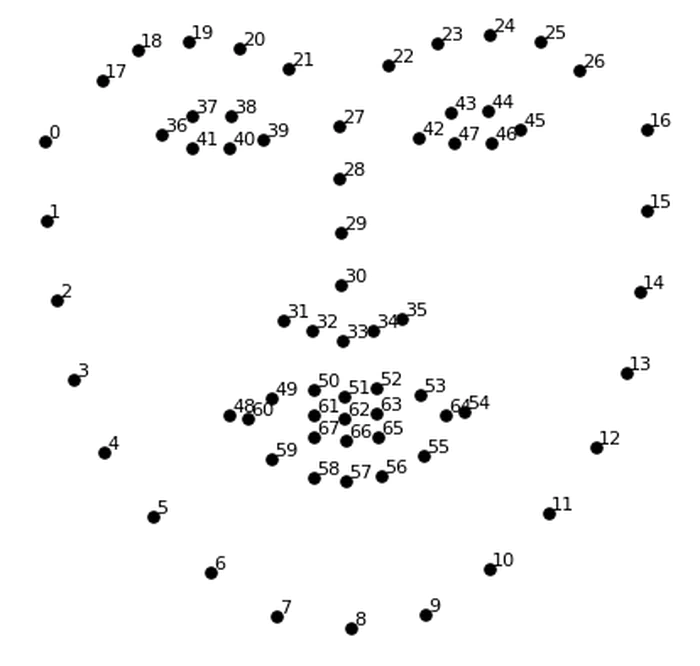}
        \caption{Facial Feature Extraction \cite{inurface}}
        \captionsetup{justification=centering}
        \label{fig:ffe}
    \end{figure}
    
     \begin{figure}[!tbp]
  \centering
  \begin{minipage}[b]{0.2\textwidth}
    \includegraphics[width=\textwidth]{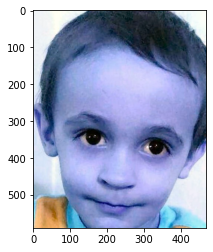}
    \caption{Before Facial Landmark Extraction}
    \captionsetup{justification=centering}
    \label{fig:beforeffe}
  \end{minipage}
  \hfill
  \begin{minipage}[b]{0.2\textwidth}
    \includegraphics[width=\textwidth]{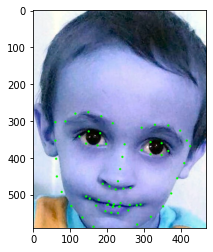}
    \caption{After Facial Landmark Extraction}
    \captionsetup{justification=centering}
     \label{fig:afterffe}
  \end{minipage}
\end{figure}
    
    For the sake of extracting features from the images of ASD and non-ASD individuals in the dataset used \ref{fac}, or in other words the facial landmarks, such as the eyes, nose, mouth, and brows, we use Python's open source library built for face recognition, Dlib, \footnote{http://dlib.net/}, proposed by \cite{dlib}. This library works by detecting a face within the input image. Then, within each face detected, it recognizes the facial landmarks and specifies their coordinates depending on a pre-defined facial template represented in Figure \ref{fig:ffe}. Thus, each point in the face would belong to a part of the face and have a unique number between 0 and 67, as well as coordinates (x, y).  In Figure \ref{fig:ffe}, the facial template, points 0 through 16 represent the jaw points, the right and left brows are represented by points 17–21 and 22–26 respectively, the points 27–35 constitute the nose, 36–41 and 42–47 belong to the right and left eyes, 48–60 belong to the mouth points, and the points range 61–67 represent the lips. Therefore, to extract facial distances, representing facial features for each facial image in hand, we use Python Dlib to develop a program that takes as input the facial image dataset \ref{fac}. Then, for each image of an individual's face in the dataset, such as the ones in Figure \ref{fig:beforeffe}, the face is recognized, and then facial landmarks, including the brows, eyes, mouth, and nose, are distinguished through the plotted dots in Figure \ref{fig:afterffe}. Later, each plotted dot in the figure is characterized by a number in the range of 0-67, an abscissa "x," and ordinate "y" in the 2D plane. For the sake of calculating facial distances from the recognized dots of the face, we take the midpoints of each facial entity depending on the pre-defined Dlib template shown in Figure \ref{fig:ffe}, and compute the euclidean distance between them. To clarify, we take the midpoint of the left brow and that of the right brow, which are points A(19) having coordinates ($x_1$, $y_1$) and B(24) having coordinates ($x_2$, $y_2$) respectively. We then evaluate the euclidean distance between points A and B, depending on equation \ref{euc}:
    
    \begin{equation} \label{euc}
        \sqrt{(x_2 - x_1)^2 + (y_2 - y_1)^2 }
    \end{equation}
    
   Finally, and similarly to the euclidean distance example calculated between points A and B, the euclidean distances are calculated between the right and left brows, the right and left eyes, and the nose and lips. Then, three facial distances are obtained for each of the patients' faces in the facial image dataset in hand. This step enhances the ability to merge the image dataset by adding the three distances, as features, to the behavioral CSV data. 
    
    \vspace{3mm}
    
    \item \textbf{Combined Data} \\
   After running our program utilizing Python Dlib on the facial image dataset \ref{fac}, three facial distances are evaluated for each facial image in the dataset. Therefore, after evaluation, these distances are added as features on top of the behavioral data features constituting the initial behavioral dataset. Then, our merged data ended up containing 22 features, 19 for the behavioral data and the remaining 3 for the facial distances. For the sake of testing in this experiment, data instances (rows) are merged randomly. To the best of our knowledge, no dataset combining both behavioral and facial features of ASD and non-ASD individuals has been found in the literature. For example, the first behavioral instance belonging to an ASD patient is mapped to the first ASD facial distance instance. This process is repeated till all of the instances are in hand. Due to the blurriness of some facial images, the Dlib program was unable to recognize some facial landmarks, which is why we ended up with 2677 data instances rather than 2940, the original dataset size.
    
\end{enumerate}


\section{\textbf{Experiments and Results}} \label{results}
In this section, we first introduce the datasets used for training our federated learning models. Second, we discuss each of the latter and give an overview of the federated learning framework used for accomplishing our experiments. Third, we illustrate the experimental results obtained.

\subsection{\textbf{Datasets}} \label{datasets}
Two datasets were used in our experiments as our work is based on two aspects of Autism Spectrum Disorder (ASD). Each of these datasets is described in the sequel.
    
   \vspace{3mm}
    \begin{enumerate}
        \item \textbf{Autism Screening Behavioral Data \cite{behvdataset}} \label{behv}
        
        \begin{itemize}
        \item \textit{\textbf{Data Description}}
        
        This dataset is a collection of questionnaire-response instances tackling the behavioral traits and aspects of ASD \cite{behvdataset}. It is collected from responses to a survey on ASD. Patients or their parents answered certain behavior-related questions, in addition to some personal details, including age, gender, country, and ethnicity. These responses are recorded in a dataset of 21 columns, representing the features, and 705 rows, corresponding to the responses. 
        
        \item \textit{\textbf{Data Pre-Processing}}
        
       In order to avoid bias in the training stage, the dataset is cleaned and preprocessed. First, rows containing any missing answers are dropped, through which we end up with 487 data instances. Then, all text answers are transformed into numbered categories so that each class of answers would correspond to a number rather than words, given that all of the features at hand are categorical rather than continuous values. For instance, in the "Gender" feature, a "Male" answer is converted to 0 and a "Female" answer is converted to 1, and so on. This method has been proven to enhance the learning ability of the ML model. Moreover, the feature "ASD", indicating whether the responder is autistic or not, or the class feature, is also categorical, so a "Yes" answer is represented by 0 and "No" by 1. It is worth mentioning that data is evenly split into both classes, to preclude any bias during training.
        Finally, we save the pre-processed CSV data of responses into a python pickle file for faster and easier access later on in the training process in the FedML framework.
        \end{itemize}
        
        \vspace{3mm}
        \item \textbf{Autism Facial Image Data \cite{facialdataset}} \label{fac}
        
        \begin{itemize}
        \item \textit{\textbf{Data Description}}
        
       The second dataset used in this paper is facial images of autistic and non-autistic patients. It consists of 1470 images of ASD patients and another 1470 images of non-ASD individuals. Hence, the data is already unbiased and split in half between the two classes. All of the images are close to the faces of patients, showing the details of their facial features, which is what we target mainly in our study. 
        
        \item \textit{\textbf{Data Pre-Processing}}
        
To prepare the dataset for training, the data undergoes several preprocessing steps. First, we convert the images into grayscale using Python, for better learning of the facial features themselves. Plus, each of the images is then converted into a multi-dimensional array, which is fed to the federated learning model. The latter could then take as input an array corresponding to the image of a patient, and then return whether they have ASD or not based on the feature values, symmetric to their facial characteristics, within the input array. In the end, image array data is also saved into a python pickle file for the federated learning model to access and load faster and easier during the training process. 
        
    \end{itemize}
    \end{enumerate}

\subsection{\textbf{Setup}} \label{exps}
In this section, we tackle the framework used for experimenting with ASD prediction using our federated learning scheme. We also describe the ML models used in the testing phase.  
    \begin{enumerate}
        \item \textbf{Federated Learning Framework}:  \label{frame}\\
        To conduct our experiments, we use the LocalFed framework \footnote{https://github.com/arafeh94/localfed} built on top of FedML \cite{fedml} for imitating the federated learning context. This is an easy-to-use and editable adaptive framework, built with all federated learning components. The major components of this system are summarized in Table \ref{tab:1}.

\begin{table*}[t]
  \begin{tabular} {|p{3cm}|p{14cm}|}
  \hline
\noalign{\smallskip}
\centering \textbf{Entity} & \textbf{Description} \\ [0.3cm]

\hline
\noalign{\smallskip}
\centering
    Trainer Manager & An instance of this interface defines how trainers are running. Trainer Manager (TM) is followed by certain training parameters defined for each client, such as epochs and loss. 
    \\ [0.3cm]
\hline
\noalign{\smallskip}
    \centering Aggregator & Defines how would the local models be merged into one global model, or in other words, aggregated. 
    \\ [0.3cm]
\hline
\noalign{\smallskip}
    \centering Client Selector & Is an interface responsible of controlling the clients selected to train in each round. 
    \\ [0.3cm]
\hline
\noalign{\smallskip}
    \centering Metrics & Specify the accuracy metrics used to evaluate the model on test data at each round. \\
\hline
\noalign{\smallskip}
    \centering Model & Defines the ML model used for training, such as logistic regression. \\ 
\hline
\noalign{\smallskip}
    \centering Number of Rounds & An indicator for how many rounds would the federated learning task run. \\ 
\hline
\noalign{\smallskip}
    \centering Data Loader & An instance of this entity enables loading and distributing the data to a number of specified clients. \\
    \hline 
  \end{tabular} \\
  \caption{Experimental Simulation Settings}
  \captionsetup{justification=centering}
  \label{tab:1}
\end{table*}

\vspace{1mm}
    \item \textbf{Machine Learning Models}: \\
   To train the localFed framework, we must specify the ML models used for local training for each client. 
    \begin{itemize}
    \vspace{1mm}
        \item \textit{Logistic Regression:} Previous work in literature \cite{comp}, especially in binary classification, has proved the efficiency of this ML model. It is based on modelling the probability of a discrete outcome, in our case, either class ASD or non-ASD, given input variables or the features we have. This model is used for training the behavioral dataset \cite{behvdataset} given that it has multiple input features in addition to the merged data.
        
       \vspace{1mm}
        \item \textit{Neural Networks:} The neural networks model is used for classifying data in the ASD facial image dataset \cite{facialdataset}. This is a class of models commonly built with layers, inspired by the biological human brain and nervous system, which explains their name. A neural network model consists of an input layer, hidden layers, and output layer, where data is forwarded from one layer to the next one till reaching the desired accuracy. The main applications of this model, in which it has been proved highly efficient, are speech recognition, image recognition, and image classification, which is why we chose it to train on the facial image dataset \cite{facialdataset}.
        
        \vspace{1mm}
        \item \textit{Decision Trees:}
This is a supervised machine learning classifier \cite{dectree}. A decision tree is directed and rooted, with two or more outgoing edges representing possible events, or if-then rules, that may occur depending on a certain discrete function. Leaf nodes represent the possible classes that may be predicted for each data instance, which are ASD or non-ASD in our case. For each new data instance, the class is predicted based on the pathway taken through the built decision tree. This ML model is used for experimenting with our merged ASD data. 
        
         \vspace{1mm}
        \item \textit{K-Nearest Neighbors:}
The KNN classifier is a non-parametric machine learning model used in order to classify unlabeled instances by mapping them to the class of the most similar labelled instances within the dataset at hand \cite{knn}. It thus classifies the data instance through predicting a class, in our case ASD or non-ASD, based on the classes of the K-nearest, neighbors or instances, to the instance under study. We utilize this model with k = 3 neighbors to test the merged data behavior in regular ML. 
        
    \end{itemize}
    \end{enumerate}

\subsection{ \textbf{Results}} \label{res}
In this subsection, we tackle the results obtained from our federated learning scheme experiments on both datasets described earlier. We measure the accuracy as a function of the number of clients available and the maximum data size trained locally. All experiments displayed are tested on 20 epochs while varying the number of epochs does not play a role in changing the results.
    
    \vspace{3mm}
    \begin{enumerate}
    \item \textbf{Our Approach with Autism Screening Behavioral Data}
        
   Experimental results obtained when ASD screening behavioral data is trained with FedML are represented in figure \ref{fig:my_labelbeh}. In the given plot, we represent the accuracy (right y-axis) by the line as a function of the number of clients in each experiment (x-axis) and the size of data trained locally by each client (left y-axis). 
As could be seen clearly, the accuracy increases with increasing the size of data per client, achieved by decreasing the number of clients per experiment. For instance, we first test with a number of clients C = 50 clients. After the distribution of data by the localFed framework, every client receives around 20 data instances, hence a 0.48 accuracy is achieved. When the number of clients is decreased to C = 10, then the maximum data size per client is increased to around 45 data instances per client, and the accuracy increases to 0.51. The highest accuracy, equal to 0.7, is achieved with C = 3 clients and 250 data instances per client. 
    
    \begin{figure}[h]
        \centering
        \includegraphics[scale=0.6]{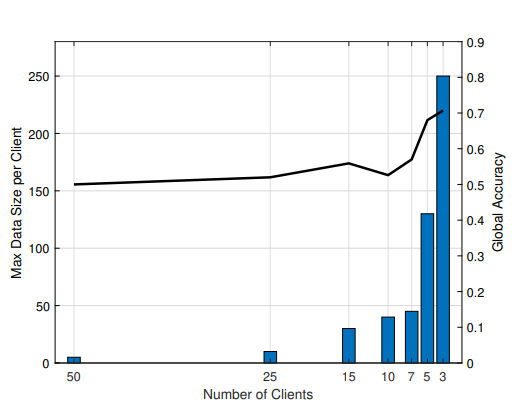}
        \caption{Results of FedML on Behavioral Dataset}
        \captionsetup{justification=centering}
        \label{fig:my_labelbeh}
    \end{figure}
    
\vspace{3mm}
    \item \textbf{Our Approach with Facial Image Data}
   
    \vspace{1mm}
   We plot, in figure \ref{fig:my_labelimg}, the experimental results noted when ASD facial image data is trained with FedML. Similar to what we did previously with behavioral data, we represent the accuracy (right y-axis) by the line as a function of the number of clients in each experiment (x-axis), with the size of the data trained locally by each client (left y-axis). 
    
    \vspace{1mm}
By examining the given plot, we see that the global accuracy increases with a higher size of data per client and decreases the number of clients per experiment. First, we experiment with a number of clients, C = 50 clients. After the distribution of data by the localFed framework, every client receives around 58 data instances, representing facial images of autistic and non-autistic individuals. An accuracy of 0.5 is achieved. When we decrease the number of clients to C = 10, then the maximum data size per client is increased to around 290 data instances per client, and the accuracy increases to 0.53. Finally, as we decrease the number of clients to C = 3 clients and simultaneously increase the maximum data size to 250 data instances per client, the highest accuracy is achieved, equal to 0.6. 
    
      \begin{figure}[h]
        \centering
        \includegraphics[scale = 0.5]{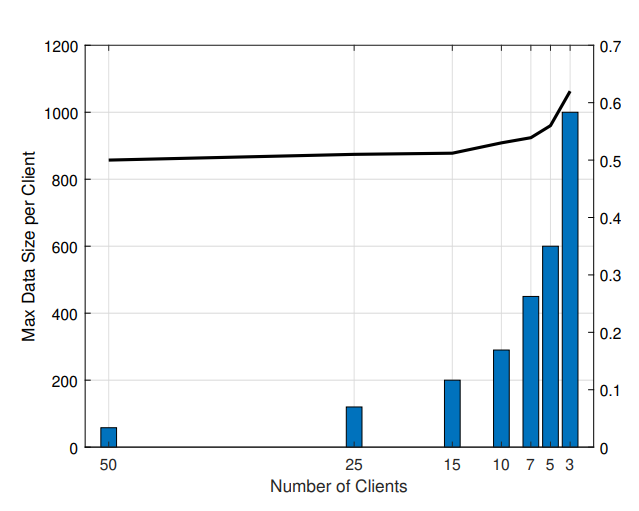}
        \caption{Results of FedML on Facial Image Dataset}
        \captionsetup{justification=centering}
        \label{fig:my_labelimg}
    \end{figure}
    
    \vspace{3mm}
   \item \textbf{Regular Machine Learning with Merged Data} 
   
   \vspace{1mm}
   
   \begin{enumerate}
    \item \textbf{\textit{Train-Test Split}}
       
    \vspace{1mm}
    In Figure \ref{fig:my_labelregm}, we plot the accuracy results obtained from testing several machine learning models on the merged data we have combined, as described in the previous section. The purpose of this experiment is to test the behavior of this data on regular machine learning models, to check which model provides the highest accuracy, before testing on our federated learning framework. For this machine learning training process, data is split regularly into 80\% training set and 20\% testing set, for the sake of measuring how far predictions are made later on from the test instances, or the prediction error \cite{sensitivity} in other words. As we can visualize from the graph in Figure \ref{fig:my_labelregm}, the highest accuracy of 0.65 is obtained with the logistic regression model. Lower accuracy values are obtained with decision trees, K-nearest neighbors with k = 3, and neural network models, which are equal to 0.62, 0.58, and 0.5 respectively. 
       
    \begin{figure}[h]
        \centering
        \includegraphics[scale=0.6]{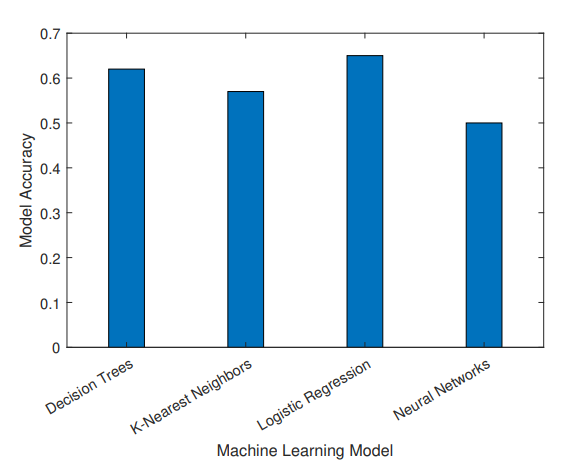}
        \caption{Regular ML with train-test split on Merged Data}
        \captionsetup{justification=centering}
        \label{fig:my_labelregm}
    \end{figure}
       
       \vspace{3mm}
    \item\textbf{\textit{K-fold Cross-Validation}}
    
   To experiment more with regular machine learning behavior on the merged data, we split the data using K-fold cross-validation (CV) instead of the usual train-test split. In K-fold cross-validation, as mentioned in \cite{crossvalid}, data is partitioned into K separate equal-sized segments, called folds. Simultaneously, training occurs on K iterations, where at each iteration one fold is selected as a testing set and the remaining K-1 folds are used for training. We use cross-validation in order to test whether higher accuracy values would be achieved than with a train-test split for the different machine learning models used. We experiment with K = 8 folds while shuffling data instances during data segmentation. As plotted in Figure \ref{fig:mergedkfold}, the accuracy results are slightly higher than those noted with the train-test split in Figure \ref{fig:my_labelregm}. The logistic regression model remains the most accurate at 0.6525, which is slightly higher than that with a regular train-test split. As for decision trees, K-nearest neighbors, and neural network models, the accuracy values noted are 0.636, 0.59, and 0.485, respectively, which are also somewhat higher than those obtained without cross-validation. 
    
    \begin{figure}[h]
    \centering
    \includegraphics[scale=0.45]{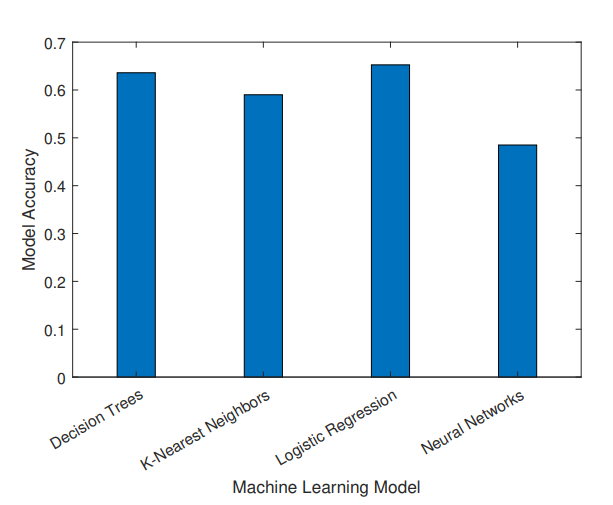}
    \caption{Regular ML on Merged Data with K-fold CV}
    \captionsetup{justification=centering}
    \label{fig:mergedkfold}
    \end{figure}

   \end{enumerate}
    
    \vspace{3mm}
   \item \textbf{Our Approach with Merged Data} 
   
   \vspace{1mm}
   After testing the merged data on regular machine learning, the best accuracy is obtained with a logistic regression model with an accuracy value of 65\%. Thus, we apply this model in our federated learning framework described in Section \ref{exps}. The obtained results, constituting the global accuracy as a function of the number of clients and data size per client, are plotted in Figure \ref{fig:my_labelfedm}. 
    
As presented in Figure \ref{fig:my_labelfedm}, the global accuracy, computed in the global aggregator of the federated learning environment, is plotted while varying the number of clients or local trainers and the size of data per client. The first experiment was done with C = 50 clients, and the data size was 53 instances per client. The global accuracy is 0.5. As we decrease the number of clients to C = 10, the data size per client thus increases to 270 data instances per client, and the global accuracy also rises to 0.52. The latter keeps rising, while simultaneously increasing the number of data instances per client and decreasing the number of training clients. Then, the highest global accuracy of 0.63 is achieved with the lowest number of clients (C = 3) and the highest data size per client (1000 data instances). Therefore, the remarkable performance of our federated learning models on the different types of data would be that the number of clients is inversely proportional to the achieved global accuracy. Fewer clients lead to a higher data size per client, and then higher accuracy. 
    
    \end{enumerate}
    
\section{\textbf{Discussion}}\label{dis}
Throughout this section, we discuss and analyze the results displayed in the previous section and give interpretations of how they were obtained. We also compare our results obtained for predicting Autism Spectrum Disorder (ASD) using our federated learning scheme to previous studies in the literature predicting the same disorder using regular ML models. 

According to what we noted in the results, similar behavior is marked with training both ASD behavioral screening data and ASD facial image data on our federated learning framework and then with the merged data as well. The most notable behavior of the plotted results would be that the global accuracy increases by decreasing the number of clients in the federated learning experiment. This note would be interpreted by the fact that when we have a higher number of clients, fewer data instances are available to each client for local training. 

Therefore, as any ML training is based on the data fed to its model, having fewer clients will eventually lead to a higher number of data instances, given that data is equally divided among clients. Thus, this redirects the trainer to a higher chance of training on larger data, and then a higher accuracy of its model. 

In comparison to regular ML models, higher accuracy is achieved in \cite{multiple} for ASD screening behavioral data and Kaggle \footnote{https://www.kaggle.com/code/cihan063/autism-cnn-vgg16} for ASD facial image data. Concerning the latter, a convolutional neural networks (CNN) model was used, which achieved a 0.82 test accuracy, higher than the best accuracy we obtained with our federated learning model, equal to 0.6 with 3 clients being available. As for the work done in \cite{multiple} for regular ML on ASD behavioral screening data, their accuracy noted was 0.9 with the artificial neural networks (ANN) model, which is also higher than the accuracy achieved in our experiments, which equals to 0.7 per 3 clients in the experiment.

 \begin{figure}[h]
    \centering
    \includegraphics[scale=0.45]{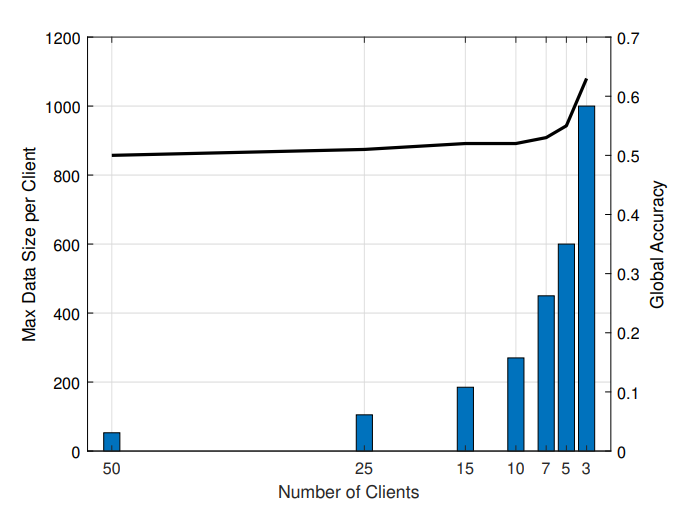}
    \caption{Results of FedML on Merged Data}
    \captionsetup{justification=centering}
    \label{fig:my_labelfedm}
\end{figure}

Surely, higher accuracy would be observed with regular ML in comparison to federated models, due to the distribution of data among clients. However, a higher level of data security and privacy is achieved when it comes to federated learning, which lacks regular ML training. This is essential in the application we are tackling, ASD prediction, given that the data being used is medical and personal, thus confidential within the screening center, and sensitive to leakage. 

Furthermore, when it comes to merged data, higher accuracy is achieved with regular ML in comparison to federated ML, due to the distribution of data instances among local trainers, as also observed with the other types of data. Moreover, the highest accuracy achieved with regular ML and federated learning on the merged data is around 65\% and 63\% respectively. Given that the merged data, through which we conducted these experiments, is randomly generated and combined, these accuracy values could be acceptable. On the other hand, higher accuracy values would be achieved if the data, combining both behavioral and facial characteristics of ASD as well as non-ASD individuals, was real rather than randomly generated and mapped. Due to the fact that we are merging two datasets, we are adding additional meaningful features to the data. This should enhance the performance of the machine learner as it would make the data/labels more detailed and specific, which improves the learning accuracy. Hence, having real ASD/Non-ASD data containing behavioral data as well as facial distances, representing real patients, rather than randomly generated and correlated, would enhance the global accuracy obtained. Moreover, the machine learner's performance would be enhanced with multi-aspect data given the medical proof we base our work on \cite{craniofacial}. Then, measuring the facial distances, the process we did through feature extraction, would be a major signal for differentiating ASD from non-ASD individuals. Therefore, the facial distance features we have added to the behavioral data should positively impact the learning process of a machine learner, especially if the facial images were real and directly related to the behavioral data instances. We would thus suggest having a dataset of facial images, with each image having a brief description of the ASD patient or normal individual, which would help more in enhancing the merging criteria for both behavioral and facial datasets. Hence, in brief, having real ASD/Non-ASD data containing behavioral data as well as facial details that belong to the same individuals, representing real patients, rather than randomly generated and correlated, would enhance the global accuracy obtained.

\section{\textbf{Conclusion}}\label{conc}
In this paper, we introduce a new federated learning scheme for predicting Autism Spectrum Disorder (ASD), characterized by special behavioral and social traits and medically proven to affect the facial traits of its patients. For this sake, we propose a prediction framework trained on two forms of data: the first consists of ASD screening behavioral data, made up of filled surveys on ASD patients' social and personal traits, and the second consists of images corresponding to ASD as well as non-ASD individuals' faces. Thus, given one of these two input datasets, our federated learning scheme would predict whether the individual is autistic or not, in a private prediction framework. We obtain remarkable results in comparison to previous regular ML models. Within our private training environment, a 71\% accuracy was obtained for ASD prediction given behavioral data, while facial image ASD data produced a 62\% accuracy. However, a higher number of data instances for both datasets in the learning process would have been more promising. For the sake of testing ASD data containing both behavioral and facial features, we tested a merged dataset via facial feature extraction using Python Dlib. Then, we study the behavior of the merged data on regular as well as federated machine learning, achieving a 65\% and 63\% accuracy, respectively. Therefore, we propose, as a future enhancement, testing the different models on real-world data instead of experimenting with synthetic data instances. We would also test the feasibility of using Federated Transfer Learning for the prediction of ASD as well as other medical conditions according to special recognizable characteristics and features of individuals.  

\section*{Acknowledgements}
This research was jointly supported by the College of Technological Innovation (CTI), Zayed University (ZU), under grant number RIF-20130, and the Lebanese American University (LAU).

\bibliographystyle{ieeetr}
\bibliography{references}

\begin{IEEEbiography}[{\includegraphics[width=1in,height=1.25in,clip,keepaspectratio]{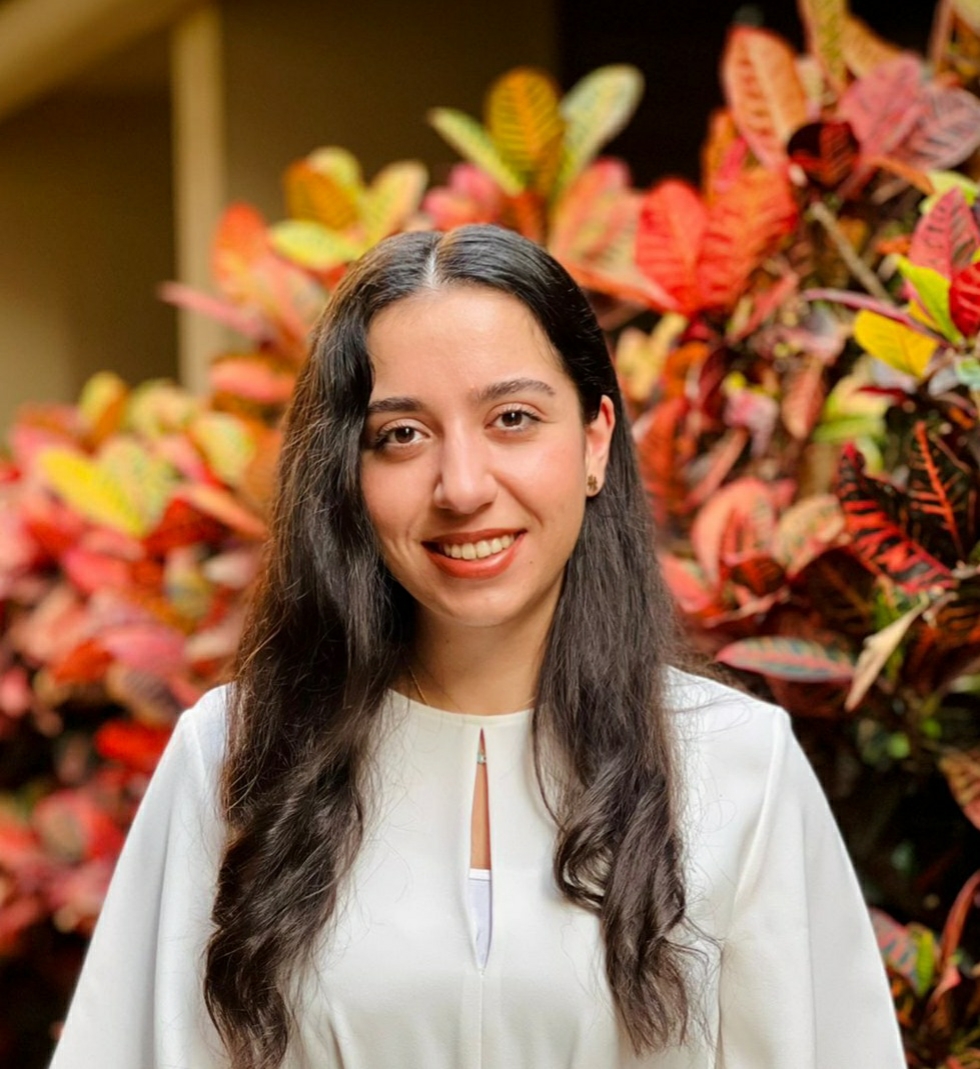}}]%
{Hala Shamseddine}
graduated with Bachelor of Science and Master of Science degrees in Computer Science from the Lebanese American University (LAU) based in Beirut, Lebanon, in June 2020 and August 2022 respectively. She is currently expanding her technical expertise within the tech field. Her research interests include fog federation formation, fog and cloud computing, federated machine learning, artificial intelligence, privacy and security.
\end{IEEEbiography}

\begin{IEEEbiography}[{\includegraphics[width=1in,height=1.25in,clip,keepaspectratio]{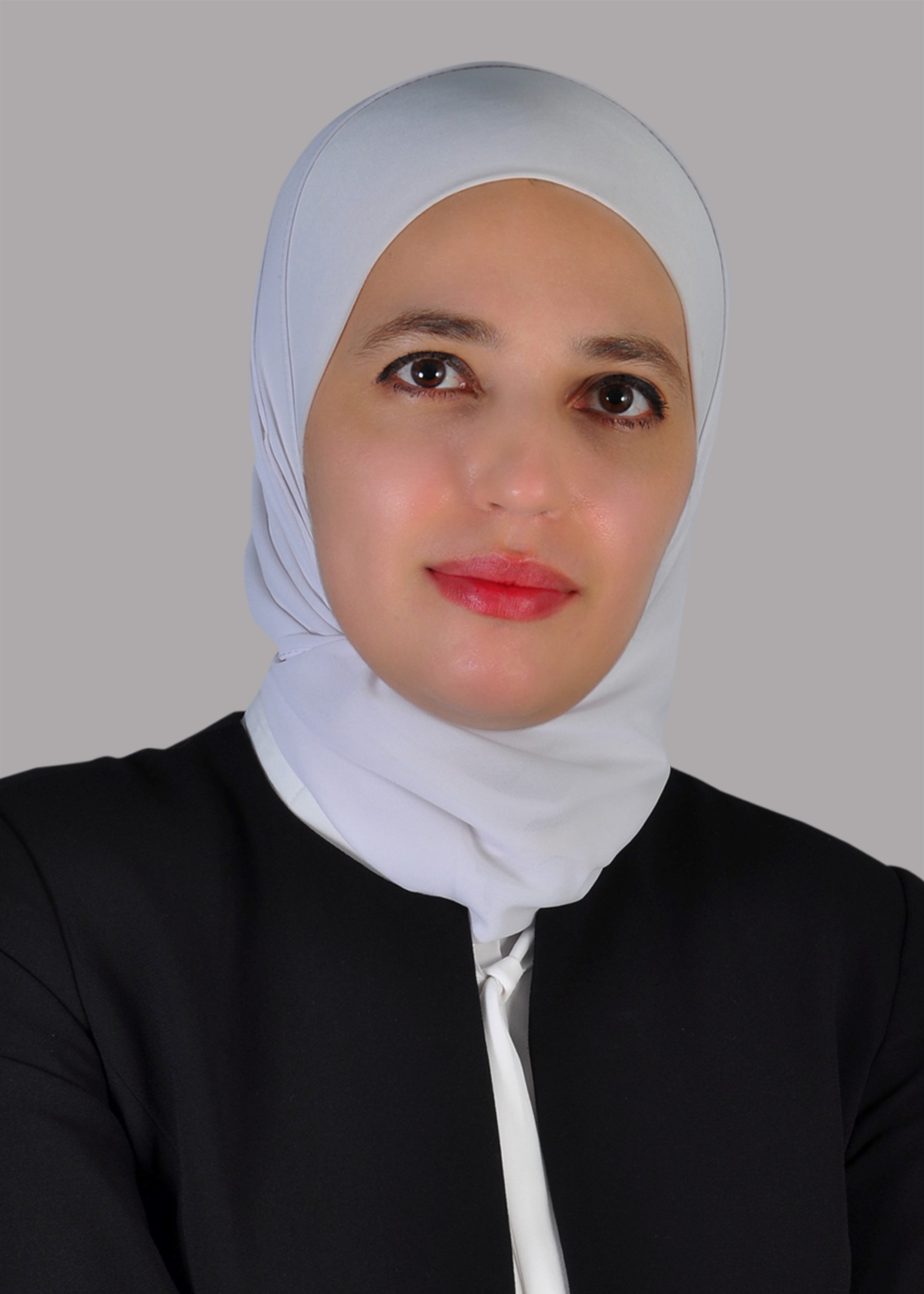}}]
{Safa Otoum}
(M’19) is an assistant professor of computer engineering in the College of Technological Innovation (CTI), Zayed University, United Arab Emirates. She received her M.A.Sc. and Ph.D. degrees in computer engineering from the University of Ottawa, Canada, in 2015 and 2019, respectively. Her research interests include blockchain applications, applications of ML and AI, IoT, and intrusion detection and prevention systems. She is a registered Professional Engineer (P.Eng.) in Ontario.
\end{IEEEbiography}

\begin{IEEEbiography}[{\includegraphics[width=1in,height=1.25in,clip,keepaspectratio]{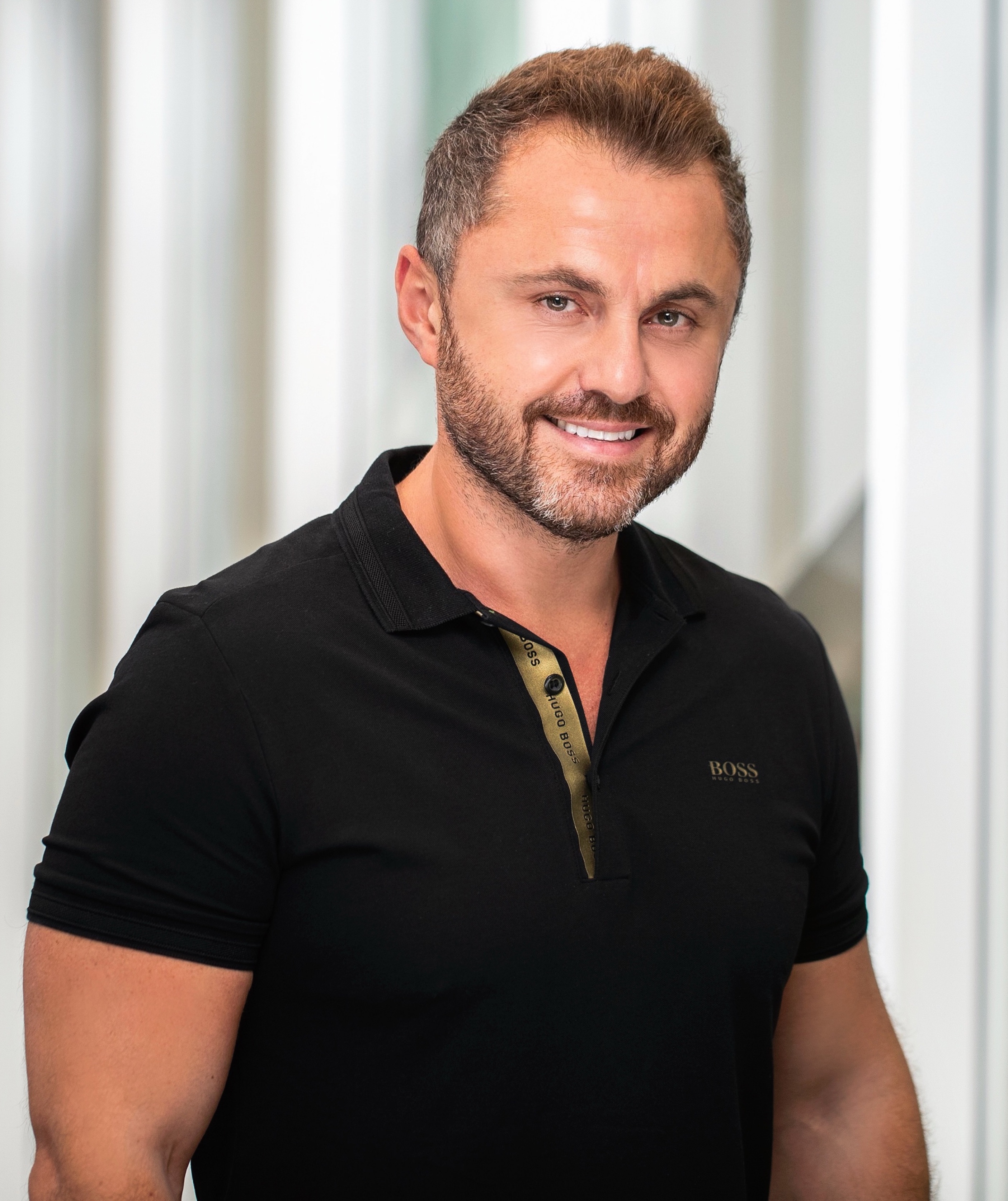}}]
{Azzam Mourad}
received his M.Sc. in CS from Laval University, Canada (2003) and Ph.D. in ECE from Concordia University, Canada (2008). He is currently Professor of Computer Science and Founding Director of the Cyber Security Systems and Applied AI Research Center with the Lebanese American University, Visiting Professor of Computer Science with New York University Abu Dhabi and Affiliate Professor with the Software Engineering and IT Department, Ecole de Technologie Superieure (ETS), Montreal, Canada. His research interests include Cyber Security, Federated Machine Learning, Network and Service Optimization and Management targeting IoT and IoV, Cloud/Fog/Edge Computing, and Vehicular and Mobile Networks. He has served/serves as an associate editor for IEEE Transactions on Services Computing, IEEE Transactions on Network and Service Management, IEEE Network, IEEE Open Journal of the Communications Society, IET Quantum Communication, and IEEE Communications Letters, the General Chair of IWCMC2020, the General Co-Chair of WiMob2016, and the Track Chair, a TPC member, and a reviewer for several prestigious journals and conferences. He is an IEEE senior member.
\end{IEEEbiography}

\end{document}